\documentclass[twocolumn]{aastex62}
\usepackage{balance}
\usepackage{placeins}
\usepackage{amsmath}
\usepackage{float}

\graphicspath{{./}{figures/}}

\shorttitle{Star Formation in Ultra-Compact Galaxies}
\shortauthors{Petter et al.}

\received{2020 June 1}
\revised{2020 August 14}
\accepted{2020 August 21 to ApJ}

\begin{document}

\title{Deviations from the Infrared-Radio Correlation in Massive, Ultra-compact Starburst Galaxies}

\correspondingauthor{Grayson Petter}
\email{Grayson.C.Petter.GR@dartmouth.edu}

\author{Grayson C. Petter}
\affil{Department of Physics and Astronomy, Dartmouth College, 6127 Wilder Laboratory, Hanover, NH 03755, USA}


\author{Amanda A. Kepley}
\affiliation{National Radio Astronomy Observatory, 520 Edgemont Road, Charlottesville, VA 22903-2475, USA}

\author{Ryan C. Hickox}
\affiliation{Department of Physics and Astronomy, Dartmouth College, 6127 Wilder Laboratory, Hanover, NH 03755, USA}

\author{Gregory H. Rudnick}
\affiliation{Department of Physics and Astronomy, University of Kansas, 1251 Wescoe Hall Dr., Lawrence, KS 66045, USA}

\author{Christy A. Tremonti}
\affiliation{Department of Astronomy, University of Wisconsin, 475 N Charter St, Madison, WI 53706, USA}

\author{Aleksandar M. Diamond-Stanic}
\affiliation{Department of Physics and Astronomy, Bates College, 44 Campus Ave, Carnegie Science Hall, Lewiston, ME 04240, USA}

\author{James E. Geach}
\affiliation{Centre for Astrophysics Research, School of Physics, Astronomy \& Mathematics, University of Hertfordshire, Hatfield, AL10 9AB, UK}

\author{Alison L. Coil}
\affiliation{Center for Astrophysics and Space Sciences, Department of Physics, University of California, 9500 Gilman Dr., La Jolla, CA 92093, USA}

\author{Paul H. Sell}
\affiliation{Department of Astronomy, University of Florida, 211 Bryant Space Science Center, Gainesville, FL 32611, USA}

\author{John Moustakas}
\affiliation{Department of Physics and Astronomy, Siena College, 515 Loudon Road, Loudonville, NY 12211, USA}

\author{David S. N. Rupke}
\affiliation{Department of Physics, Rhodes College, Memphis, TN 38112, USA}

\author{Serena Perrotta}
\affiliation{Center for Astrophysics and Space Sciences, Department of Physics, University of California, 9500 Gilman Dr., La Jolla, CA 92093, USA}

\author{Kelly E. Whalen}
\affil{Department of Physics and Astronomy, Dartmouth College, 6127 Wilder Laboratory, Hanover, NH 03755, USA}

\author{Julie D. Davis}
\affil{Department of Astronomy, University of Wisconsin, 475 N Charter St, Madison, WI 53706, USA}





\begin{abstract}

Feedback through energetic outflows has emerged as a key physical process responsible for transforming star-forming galaxies into the quiescent systems observed in the local universe. To explore this process, this paper focuses on a sample of massive and compact merger remnant galaxies hosting high-velocity gaseous outflows ($|v| \gtrsim 10^{3}$~km~s$^{-1}$), found at intermediate redshift ($z \sim 0.6$). From their mid-infrared emission and compact morphologies, these galaxies are estimated to have exceptionally large star formation rate (SFR) surface densities ($\Sigma_{SFR} \sim 10^{3}$~$\mathrm{M_{\odot}}$~yr$^{-1}$~kpc$^{-2}$), approaching the Eddington limit for radiation pressure on dust grains. This suggests that star formation feedback may be driving the observed outflows. However, these SFR estimates suffer from significant uncertainties. We therefore sought an independent tracer of star formation to probe the compact starburst activity in these systems. In this paper, we present SFR estimates calculated using 1.5~GHz continuum Jansky Very Large Array observations for 19 of these galaxies. We also present updated infrared (IR) SFRs calculated from WISE survey data. We estimate SFRs from the IR to be larger than those from the radio for 16 out of 19 galaxies by a median factor of 2.5. We find that this deviation is maximized for the most compact galaxies hosting the youngest stellar populations, suggesting that compact starbursts deviate from the IR-radio correlation. We suggest that this deviation stems either from free-free absorption of synchrotron emission, a difference in the timescale over which each indicator traces star formation, or exceptionally hot IR-emitting dust in these ultra-dense galaxies.


\end{abstract}



\section{Introduction} \label{sec:intro}


A long-standing challenge in understanding the formation of galaxies in a $\Lambda$-Cold Dark Matter ($\Lambda$CDM) universe is the tendency for models to overpredict the number of both high and low mass galaxies formed by the present day when compared to observations \citep{Benson, Croton}. A favored solution to this ``overcooling'' problem is to invoke feedback from massive stars and active galactic nuclei (AGN), both of which can heat and expel gas from galaxies, reducing their ability to form new stars. At the massive end of the halo mass function, it has been suggested that these outflows are dominantly driven by AGN feedback, as the mass of a black hole is directly proportional to the mass of the hosting galactic bulge \citep[e.g.,][]{Ferr, geb, kor} and therefore roughly proportional to the mass of the host galaxy. Hence, a massive galaxy should harbor a proportionally massive central black hole capable of injecting large amounts of energy into the surroundings through AGN activity.


Alternatively, it is possible that stellar feedback plays an important role in galaxy evolution even in massive systems. Recently, a sample of starburst galaxies has been discovered at $z \sim 0.6$ hosting energetic outflows that appear to be driven by compact starburst activity. These galaxies lie on the massive end of the stellar mass function \citep[$M_{*} \sim 10^{11}$~$\mathrm{M_{\odot}}$,][]{Diamond2012}, are extremely compact \citep[$R_{e} \sim \textnormal{few} \ 100$ pc,][]{sell}, and host high-velocity gaseous outflows ($|v| \gtrsim 1000$ km~s$^{-1}$) appearing to actively quench star formation \citep{Tremonti, geach1, geach2, rupke}. IR SFR estimates from \textit{Wide-field Infrared Survey Explorer} (WISE; \citealt{wise}) 12 $\&$ 22 $\mu$m fluxes combined with physical size estimates from Hubble Space Telescope (HST) imaging yield extreme SFR surface densities \citep[$\Sigma_{SFR} \sim 10^{3}$~$\mathrm{M_{\odot}}$~$\textnormal{yr}^{-1}$~$\textnormal{kpc}^{-2}$, ][]{Diamond2012}, approaching the theoretical Eddington limit of radiation pressure on dust grains, or the self-limited maximum possible density of star formation \citep{lehnert, meurer, murray, thompson}. This suggests that photon pressure and other stellar feedback processes could be driving the outflows. \citet{sell} used multiwavelength observations to characterize AGN activity in a sub-sample of these galaxies selected to have the highest probability of hosting an AGN, but found no evidence of AGN activity in half of the sample and argued that an AGN contributes $<10\%$ of the bolometric luminosity in the remainder. These results appear to substantiate the claim that stellar feedback is responsible for powering the observed outflows.

Studying feedback processes in these extreme objects may serve as an important probe into the evolution of massive galaxies. Compact star-forming galaxies have been found to be much more common at $z > 3$ and are the likely progenitors of the compact galaxies that dominate the quiescent population at $z>1.5$, but their faintness makes them challenging to study in detail. Therefore, this population at $z \sim 0.6$ may serve as an analog to higher redshift massive galaxy populations and provide an opportunity to understand massive galaxy evolution at high redshift. A robust characterization of the SFRs of our sample galaxies to probe stellar feedback is critical to this endeavor.


It is difficult to robustly estimate SFRs for these young and compact starbursts, as they  may have significant differences in physical conditions from typical star-forming galaxies (SFGs) which cause the IR SFR estimates to be inaccurate. For instance, compact starbursts have been shown to exhibit a mid-IR to total IR (TIR/IR, $\lambda =$ 8--1000 $\mu$m) enhancement \citep{groves}; the IR is expected to overestimate the instantaneous SFR in young starbursts \citep[e.g.,][]{hayward}; and mid-IR emission from compact galaxies may be attenuated by dust extinction \citep{lutz}. In addition, the WISE W3 and W4 passbands lie at the short-wavelength tail of the galaxies' IR spectral energy distribution (SED) at the redshift of this sample, meaning a small variation in the SED shape could translate to a large change in the estimated IR luminosity. We therefore sought a star formation tracer dependent on longer wavelength emission that is optically thin to dust and less sensitive to the choice of the model. Using synchrotron continuum emission as an SFR indicator fulfills these requirements, and thus we present 1.5~GHz continuum observations of 20 of these galaxies here. These observations provide an independent probe of the SFRs and also further constrain ongoing AGN activity.

Radio continuum emission traces star formation because it is generated by the activity of massive stars in normal SFGs. At low frequencies, the radio continuum is dominated by synchrotron (non-thermal) emission, which is produced by cosmic ray electrons accelerated by supernova explosions of massive stars spiraling in the galactic magnetic field \citep{radiorev}. Thermal (free-free/bremsstrahlung) radiation at radio frequencies also traces star formation, as the ionizing photon rate (supplied by massive stars) is directly proportional to a galaxy's thermal luminosity. Other processes that can affect the radio continuum include the converses of each of the emission processes, synchrotron and free-free absorption. Finally, AGN activity can generate synchrotron emission. Therefore, we must assume that radio continuum emission is dominated by star-formation activity and is not self-absorbed in order to estimate an accurate SFR from a radio continuum observation.

For typical SFGs, there is a clear relationship between SFRs derived from the IR and radio, due to the well-established and remarkably tight IR-radio correlation \citep[IRRC, e.g.,][]{Jong, Helou} that holds over many orders of magnitude in IR and radio luminosity \citep{Yun}. This correlation is typically interpreted as a direct relationship between star formation and cosmic ray production, though the precise physical mechanism relating the two regimes has yet to be established. The prevailing explanations include the calorimetry model \citep{volk}, the optically-thin scenario \citep{hebic}, and the ``conspiracy'' model \citep{bell, lacki}. Thus, barring any differences between our sample and normal SFGs, we would expect to derive similar SFRs from IR and radio observations. Likewise, any differences between the two SFR estimates may inform us about the physical properties of the interstellar medium in these objects.

In this paper, we present 1.5~GHz continuum observations made with the NSF's Karl G. Jansky Very Large Array (JVLA/VLA) for 20 of these compact starburst galaxies. We find that only one galaxy exhibits an extreme radio luminosity characteristic of radio-loud AGN, providing another constraint on ongoing AGN activity in these systems. We also find that the IR SFR exceeds the radio SFR in the remaining sample for 16 of 19 galaxies by a median factor of 2.5, and that this deviation is maximized for the most compact galaxies with the youngest stellar populations. We conclude that this shift most likely stems either from synchrotron emission attenuation by free-free absorption in H II regions, from exceptionally hot dust in these compact star forming regions, or from a difference in the timescale over which the two SFR indicators trace star formation.


Throughout this paper, we adopt a standard $\Lambda$CDM cosmology, with H$_{0} = 70$ km s$^{-1}$ Mpc $^{-1}$, $\Omega_{\mathrm{m}} = 0.3$, and $\Omega_{\Lambda} = 0.7$. For SFR calculations, we assume a \citet{Kroupa} initial mass function (IMF). We also assume a non-thermal spectral index parameter of $\alpha^{NT} = -0.8$ at 1.5~GHz, using the convention where $S_{\nu} \propto \nu^{\alpha^{NT}}$. Finally, any magnitudes presented are in the Vega system.
\section{Observations and Data Reduction} \label{sec:datared}

We targeted 21 galaxies out of our SDSS+HST+WISE sample for radio continuum observations with the VLA. The parent SDSS sample was constructed of 159 galaxies with post-starburst features at $0.35 < z < 1$, and a portion was followed up with ground-based spectroscopy to investigate outflow characteristics of post-starbursts \citep{Tremonti}. After discovering that much of the sample was detected with WISE, 29 objects were followed up with HST imaging to investigate star-formation feedback by measuring SFR surface densities \citep{Diamond2012}. The sample in this paper is a subsample of this HST sample. Details of the construction of these samples will be presented in Tremonti et al.\ (in preparation).

We used the VLA in its highest resolution (A) configuration  to observe the 1.5~GHz ($\sim$ 2.4~GHz rest) continuum emission from 21 galaxies (VLA project numbers VLA/16B-238  and VLA/18A-127). One galaxy's image (J0901) was dynamic range limited instead of noise limited due to the presence of a bright source in the field and thus was excluded, limiting the sample further to 20 sources. We configured the correlator to provide two 512~MHz wide base bands, one centered at 1.264~GHz and the other at 1.776~GHz, each with dual polarization and 2048 channels.  The data were taken in 2016 October and 2018 March through June. The sources were observed in 30--45 min execution blocks. Each block observed a single source and included all necessary flux and bandpass calibration. The phase calibrator was observed for 1--1.5 min for every 5--6 min on source. The number of execution blocks for each source depended on the desired sensitivity. Integration times were calculated to measure 3$\sigma$ detections of the luminosity associated with an SFR of 30 M$_{\odot}$ yr$^{-1}$ according to Equation \ref{eq:sfr}.

The data were reduced using the VLA calibration pipeline (Pipeline version Pipeline-CASA5.4-P2-B and CASA version 5.4.1-32). After an initial calibration, we inspected each plot of the phase and amplitude for the calibrator sources for each execution block, and manually flagged visibilities affected by radio-frequency interference (RFI) or misbehaving antennas. This process removed most of the aberrant data. To remove the remaining RFI, we iteratively ran the CASA task \textit{tfcrop} to fit both the target and calibration data and automatically discard any visibilities lying more than three standard deviations away from the fit in frequency and four standard deviations in the time domain. Three iterations of \textit{tfcrop} removed most of the remaining RFI. A median of 22\% of the total 1.024~GHz bandwidth was lost to RFI (typical of 1.5~GHz observations), leaving us with a median effective bandwidth of $\sim$800~MHz. The loss of bandwidth was included in the sensitivity calculations for the observing proposal by using a conservative estimate of the expected effective bandwidth (600~MHz) after RFI removal instead of the nominal bandwidth provided by the correlator. The minimum effective bandwidth for our non-detected sources was $\sim$~750~MHz, ensuring that our detection capability was not limited by RFI.

The calibrated data were imaged using the \textit{tclean} task in CASA 5.3.0-143\footnote{Calibrating data in one version of CASA and imaging in another does not have an effect on the final image in this case.} \citep{Casa}. At 1.5~GHz the primary beam of the VLA is half a degree across. Imaging this large a field requires correcting for the non-coplanarity of the baselines and the curvature of the sky. To account for this, we gridded the data using w-projection with 128 planes. Although we did not expect our sources to be resolved, we used a robust weighting of 0.5 to provide high resolution as well as sensitivity to possible extended structures. The deconvolution was performed using multi-term multi-frequency synthesis with two Taylor terms and three size scales of point source scale, $\sim2\arcsec$, and $\sim6\arcsec$. The auto-multithresh algorithm within \textit{tclean} \citep{kepley} was used to automatically mask sources during the cleaning process. Our sources are all close to the center of the image, which reduces any off-axis effects related to measuring fluxes and source morphologies.

Given the substantial  number of targets, we developed a simple automated imaging pipeline to image the data. For each source, the pipeline first constructs a dirty image of each galaxy and then estimates the noise in this image using the corrected median absolute deviation (MAD), a robust noise estimator. The MAD is an appropriate noise estimator for our sparsely populated images. The threshold for our subsequent \textit{tclean} run was set to three times the estimated noise and the images were cleaned using the parameters described above. The resulting images were primary beam corrected and a small cutout ($\sim 1 \, $arcmin$^{2}$) of the image was made containing only the source of interest and its immediate surroundings to simplify analysis.

The resolution of our resulting images is $\sim 2\arcsec \ $, though beam parameters and noise levels vary across the images. Precise values of these parameters are listed in Table \ref{table:obs}. Each image probes an effective observed continuum frequency of $\sim 1.5$~GHz, which is equivalent to $\sim 2.4$~GHz rest frame for our target galaxies.

\tabcolsep=0.2cm
\begin{deluxetable}{cccccc}
\caption{Observation statistics of our VLA images.}
\tablehead{\colhead{Source} & \colhead{Noise} & \colhead{$B_{max}$\tablenotemark{a}} & \colhead{$B_{min}$\tablenotemark{b}} & \colhead{P.A.\tablenotemark{c}} & \colhead{$t_{\mathrm{int}}$\tablenotemark{d}}\\ \colhead{ } & \colhead{$\frac{\mathrm{\mu Jy}}{\mathrm{beam}}$} & \colhead{$\mathrm{{}^{\prime\prime}}$} & \colhead{$\mathrm{{}^{\prime\prime}}$} & \colhead{$\mathrm{{}^{\circ}}$}  & \colhead{hr}}
\startdata
J0106--1023 & 20.9 & 2.0 & 1.1 & -30.0 & 0.5\\
J0826+4305 & 22.0 & 2.1 & 1.1 & -72.9 & 0.5 \\
J0827+2954 & 11.7 & 1.3 & 1.0 & -73.8 & 2.1\\
J0905+5759 & 25.3 & 2.5 & 1.0 & -69.0 & 0.5\\
J0908+1039 & 18.8 & 2.0 & 1.2 & 64.6 & 0.6\\
J0944+0930 & 18.6 & 1.2 & 0.9 & -4.0 & 0.7\\
J1039+4537 & 11.0 & 2.6 & 1.1 & 68.7 & 1.5\\
J1107+0417 & 25.0 & 1.5 & 1.2 & -24.0 & 0.5\\
J1125--0145 & 19.9 & 2.1 & 1.1 & -43.1 & 0.7\\
J1219+0336 & 22.5 & 1.6 & 1.1 & -47.3 & 0.5 \\
J1229+3545 & 16.3 & 1.5 & 1.1 & -81.1 & 1.3\\
J1232+0723 & 25.5 & 1.7 & 1.3 & -45.2 & 0.5\\
J1248+0601 & 11.8 & 2.1 & 1.1 & -54.9 & 1.5\\
J1341--0321 & 15.8 & 1.5 & 1.2 & 13.6 & 1.8\\
J1506+6131 & 25.3 & 3.1 & 1.1 & -61.3 & 0.5\\
J1613+2834 & 26.3 & 1.6 & 1.2 & -86.3 & 0.5\\
J2116--0634 & 12.5 & 1.5 & 1.1 & 17.9 & 2.8\\
J2118+0017 & 24.6 & 1.3 & 1.1 & 22.0 & 0.5\\
J2140+1209 & 9.8 & 1.3 & 1.1 & 1.0 & 3.3\\
J2256+1504 & 9.1 & 1.2 & 1.1 & 29.8 & 2.8
\enddata
\label{table:obs}
\tablenotetext{a}{Major Axis of Synthesized Beam}
\tablenotetext{b}{Minor Axis of Synthesized Beam}
\tablenotetext{c}{Position Angle of Synthesized Beam}
\tablenotetext{d}{Observation Integration Time}
\end{deluxetable}

\section{Characterizing Source Properties}
\subsection{Radio Flux Measurement} \label{flux-meas}

We used the \textit{imfit} task within CASA to fit the target sources with 2D Gaussians in order to measure integrated flux densities \citep{fits, Vlass}. Since our sources were unresolved by our $\sim 2 \arcsec$ beam, we chose to fix the Gaussian major and minor widths, as well as the position angle to be the same as the synthesized beam. To account for the VLA's relatively low resolution and pointing accuracy compared with that achieved by HST, we fixed the center of the Gaussian as follows: if the value in any pixel within a 1$\arcsec$ radius aperture of the HST centroid is greater than 3 times the noise, we fixed the center of the fit to the brightest pixel in the aperture. This method was applied to all of our detected sources (see Figure \ref{fig:Stamps}) and improved the signal to noise of our detections. Otherwise, the center position was supplied by the centroid of an HST image of the same galaxy. 

The fit returns an integrated flux and a peak flux, both with uncertainties. Here, we adopt the detection criteria for Gaussian fits to unresolved sources used in \cite{Vlass}. A source is thus deemed a detection if the geometric mean of the peak and the integrated flux is greater than 3 standard deviations above the image noise. 

For detections, we could now use the flux and uncertainty to calculate the SFR, as described in the following section. For non-detections, we set an upper limit on the flux to be 3 times the image noise, which could then correspondingly be converted into an upper limit on the SFR. 

\begin{figure}
\includegraphics[width=0.45\textwidth]{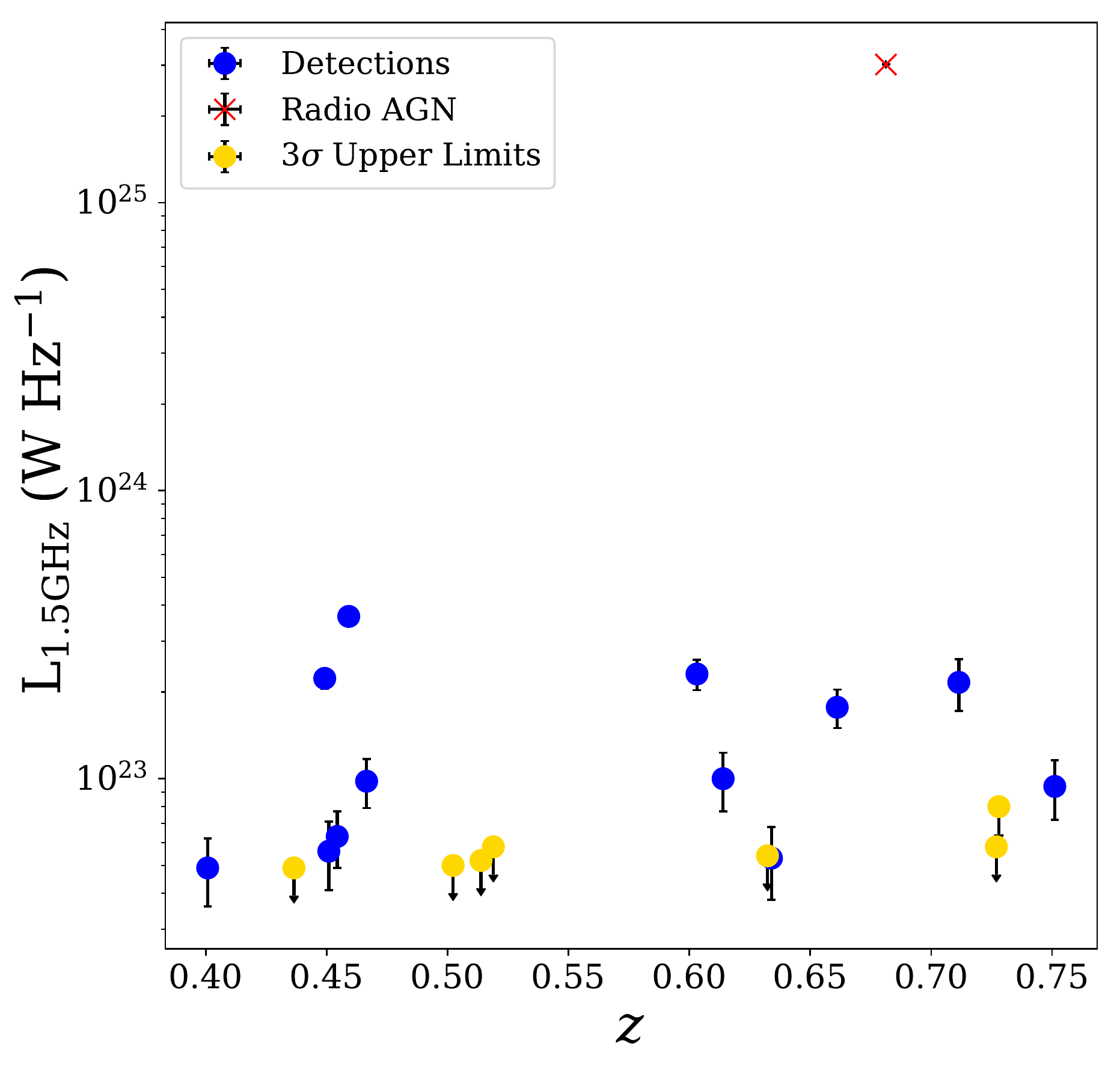}
\caption{The 1.5~GHz luminosity versus redshift for each of our sources. We show the detections in blue, the non-detections located at their $3 \sigma$ upper limits in yellow, and the one radio-loud AGN denoted by a red cross. }
\label{fig:lum_z}
\end{figure}

With our 1.5~GHz JVLA observations we detect 13 sources and have 7 upper limits. Figure \ref{fig:lum_z} demonstrates that all of our non-detections are the lowest luminosity objects and that there is only a weak trend of detection capability with redshift. The upper limit luminosity varies by less than a factor of two over the redshift range $0.4 < z < 0.75$.

\subsection{Radio-Loud AGN} \label{sec:sample}

Figure \ref{fig:lum_z} demonstrates that one of the sources (J0827) in our VLA sample has an extreme radio luminosity, and visual inspection of the radio image reveals jet structure characteristic of a radio-loud AGN. This source's radio luminosity far exceeds the $3 \sigma$ radio excess criterion used in \citet{smolcic} to diagnose AGN-dominated radio sources. As this study aims to measure SFRs of our sample galaxies through their radio continuum emission, we thus exclude this AGN-dominated source from the remainder of the analysis except where otherwise noted. 

\subsection{Radio Star Formation Rate Calculation}

We used the following equation from \citet{murphy11} to calculate the radio SFRs:
\begin{equation} 
\begin{aligned}
\bigg(\frac{\mathrm{SFR}_{\nu}^{NT}}{\mathrm{M_\odot {yr}}^{-1}}\bigg) & = 6.64\times 10^{-22}\bigg(\frac{\nu}{\textnormal{GHz}}\bigg)^{-\alpha^{NT}} \bigg(\frac{L_{\nu}^{NT}}{\textnormal{W} \  \textnormal{Hz}^{-1}}\bigg).\\ \\
\end{aligned}
\label{eq:sfr}
\end{equation}

This  equation is calibrated for the calculation of the SFR of a galaxy given a non-thermal spectral luminosity $L_{\nu}^{NT}$ of a galaxy observed at a frequency $\nu$. It assumes a \citet{Kroupa} IMF down to $ 0.1 \ \mathrm{M_{\odot}}$, and a supernova cutoff mass of $8 \ \mathrm{M_{\odot}}$. The spectral index parameter in Eq. \ref{eq:sfr} is of the opposite sign here as found in \citet{murphy11}, as here we adopt the convention where $S_{\nu} \propto \nu^{\alpha^{NT}}$. We adopt the non-thermal spectral index parameter of $\alpha^{NT} = -0.8$ given in \citet{radiorev} as the index at the rest frame frequency of our sample ($\sim 2.4$~GHz) as well as in the observed frame ($1.5$~GHz). We note that varying the spectral index parameter by 10\% of our chosen value of $\alpha^{NT} = -0.8$ yields a corresponding systematic uncertainty on the radio SFRs of $\sim 6\%$.

Although synchrotron emission dominates the radio continuum at GHz frequencies, we allow for a small thermal component ($\sim 10\%$) to the observed continuum luminosity according to the non-thermal to thermal ratio given in \citet{taba}. We note that this assumes that galaxies at intermediate redshift exhibit similar non-thermal fractions as local galaxies. We thus estimate the non-thermal luminosity as a fraction of the observed continuum luminosity $L_{\nu}$ as:

\begin{equation} 
\begin{aligned}
L_{\nu}^{NT} = \left( 1 +  \frac{1}{13} \left( \frac{(1+z)\nu }{\mathrm{GHz}}\right)^{-0.1-\alpha^{NT}} \right)^{-1} L_{\nu}.
\end{aligned}
\label{eq:ntfrac}
\end{equation}

To calculate a spectral luminosity applicable to Eq. \ref{eq:sfr} from a flux density measured in Section \ref{flux-meas}, we recall the familiar inverse square law:
\begin{equation} 
F_{\nu} = \frac{L_{\nu}}{4\pi D_{L_{\nu}}^2},
\label{eq:inv}
\end{equation}
using the spectral luminosity distance $D_{L_{\nu}}$, as defined by combining 3 equations given in \cite{condon} as:
\begin{equation} 
D_{L_{\nu}}^{2} =D_{C}^{2} (1+z)^{1-\alpha},
\label{eq:Dlv}
\end{equation}
where $D_C$ is the comoving distance to the source. Here, instead of evaluating the comoving distance integral numerically, we employ the polynomial approximation given in \cite{condon}.

By combining equations \ref{eq:sfr}--\ref{eq:Dlv}, we can derive an equation for the radio continuum SFR which depends on the observed flux density $F_{\nu}$ from a source at redshift $z$. 


\subsection{Infrared Star Formation Rate Calculation}

In addition to new estimates of the SFRs from radio data, we also present SFRs calculated from mid-IR data that are updated from those previously presented in \citet{Diamond2012}. The new SFRs are calculated using unWISE \citep{unwise} W3 and W4 data and are a median factor of 1.28 times larger than our previous estimates. At the redshift of our sample these passbands lie at the short-wavelength tail of the IR dust curve, and because SFR is calibrated on the integrated IR luminosity, a template must be used to extrapolate to the dust peak. The templates we use here are IR SEDs derived from observations of dusty IR-luminous galaxies at $0.5 < z < 3$ \citep{kirkpatrick}. We suggest these are the best available templates as we believe our sample of ultra-compact galaxies to be good analogs of galaxies at higher redshift. Here, we exclude the 4 out of 11 templates which are calibrated on observations of galaxies with powerful AGN. This choice was motivated by the lack of significant observed AGN signatures in the radio (Figure \ref{fig:lum_z}), the mid-IR (Figure \ref{fig:wisecolor}), and at other wavelengths \citep{sell}. This leaves only templates for galaxies which are star-forming or ``composite'' (including a subdominant contribution from an AGN). We calculate the IR SFRs using each template and take the median and the standard deviation of the distribution to be the estimate and uncertainty of the IR SFR, respectively. 

To calculate an IR SFR for a given galaxy, we shift the templates to the measured redshift, then numerically integrate the template over the WISE response curves to calculate a model spectral luminosity at each passband. We use the unWISE magnitudes and the redshift to compute the rest-frame mid-IR luminosities. Finally, we perform a least squares fit of the model luminosities to our observed luminosities using only one free parameter, the overall normalization of the model. This normalization parameter can then be multiplied by the template total IR luminosity to calculate an estimate of the total (8--1000~$\mu$m) IR luminosity of the galaxy. Three galaxies in the sample are not detected in W4, so in this case the normalization parameter is simply given by the ratio of the template and observed luminosities at the characteristic wavelength of the W3 filter ($\lambda = 12.082$~$\mu$m). Finally, we can estimate the IR SFR using the relation between total IR luminosity and SFR given by \citet{murphy11}:

\begin{equation}
    \bigg(\frac{\mathrm{SFR}_{\mathrm{IR}}}{\mathrm{M_\odot {yr}}^{-1}}\bigg) = 3.88\times10^{-37} \bigg(\frac{L_\mathrm{IR}}{\mathrm{W}}\bigg).
    \label{eq:IRSFR}
\end{equation}

\subsection{Ancillary Measurements} \label{sec:anc}
\textit{Galaxy compactness}: We utilized two methods of estimating the compactness of the galaxies in our sample. The first method was to use optical HST F814W images to measure a compactness parameter, defined as the ratio of nuclear ($r<$1~kpc) to the total ($r<$5~kpc) optical (background subtracted) flux. The HST images achieve sub-kpc resolution at this redshift ensuring these measurements are unbiased.

The second method was to use effective radii of the galaxies as a proxy for compactness. Effective radius should trace compactness as our sources occupy a small range in stellar mass \citep[$\log (M_{\star}/M_{\odot}) \sim 11 \pm 0.1$, ][]{Diamond2012}, meaning differences in size directly trace the stellar mass density. Effective radii were measured using the same optical HST F814W images \citep{Diamond2012, sell}. The F814W filter with $\lambda_{\mathrm{eff}} \sim 800 \ \mu$m probes $\lambda \sim 500 \ \mu$m emission at $z\sim0.6$, tracing the young stars rather than the stellar mass.

\textit{Mean stellar ages}: The mean stellar ages were computed by fitting linear combinations of C3K simple stellar population models (Conroy et al.\, in preparation) with dust attenuation to the galaxy SED (Tremonti et al.\, in preparation). The SEDs are composed of GALEX, SDSS, and WISE photometry and $R\sim2000$ spectra spanning 3600--4600 \AA.  The ages reported are the light-weighted average age of stellar populations younger than 1 Gyr, hence they characterize the time of the recent starburst, rather than the galaxies' extended star formation histories. The data, analysis and final stellar ages will be presented in Tremonti et al.\, (in preparation).

\textit{Outflow velocities}: Gaseous outflow velocities are calculated using ground-based spectroscopic observations of Mg II 2796 and 2803~\AA \ absorption lines. The spectra are continuum-normalized and shifted into the rest frame. The absorption profiles are then fit by applying between one and three Gaussian doublet models and an additional Gaussian doublet for P Cygni emission where required. The wavelength at which the outflow velocity is calculated comes from an equivalent width-weighted composite absorption profile. This profile is created using just the blueward doublet member of each fit component, and the wavelength at which this composite profile reaches 50\% of the total profile equivalent width is the wavelength at which the average outflow velocity is determined. Details of these computations and final velocity measurements will be presented in Davis et al.\ (in preparation).

\section{Results} \label{sec:results}

\subsection{Radio vs. IR SFR Comparison}

\begin{figure}[h]
\includegraphics[width=0.47\textwidth]{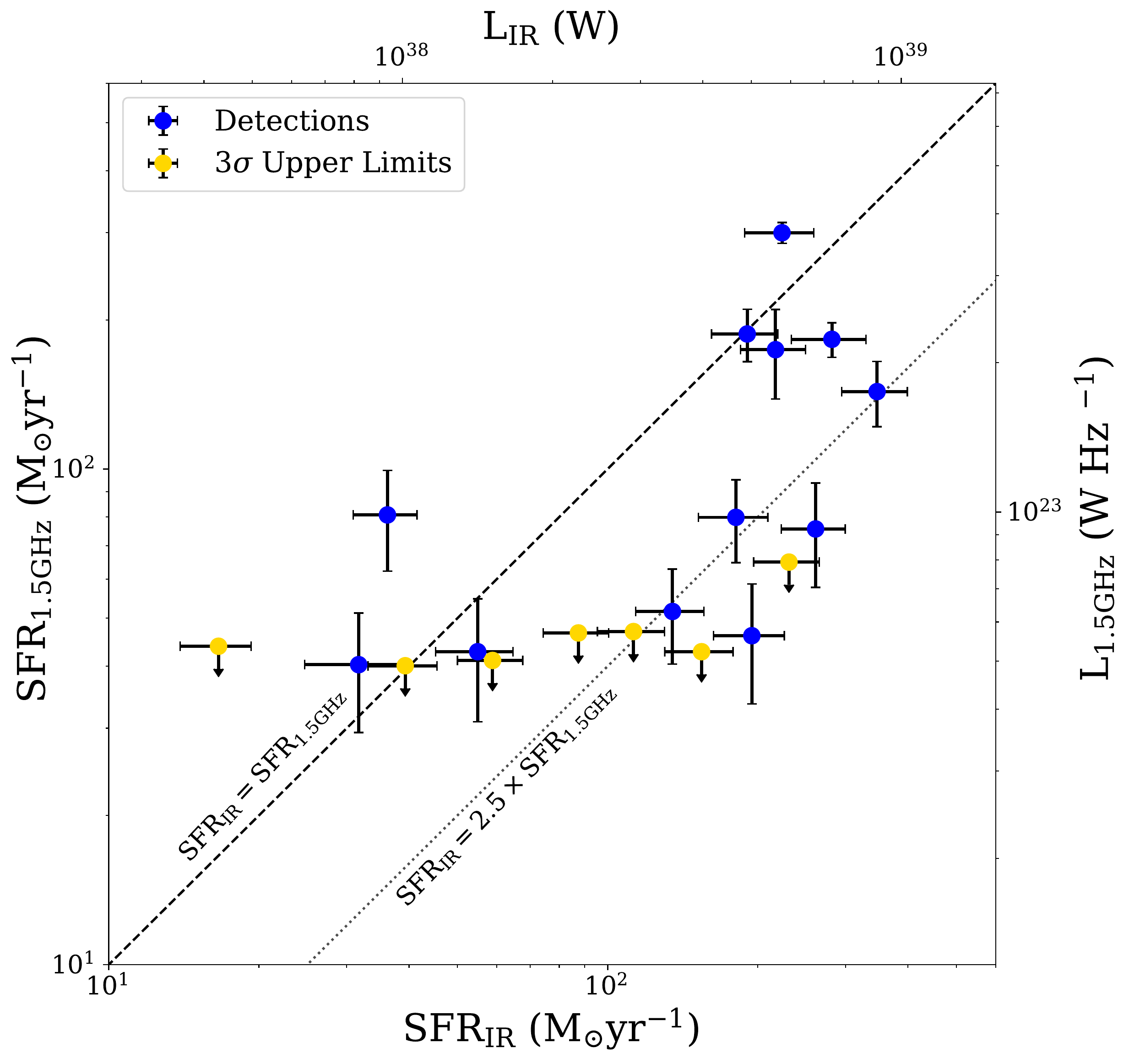}
\caption{A comparison between the SFRs as estimated by the radio and IR luminosities. We show the radio detections in blue and the non-detections in yellow. Secondary axes on the top and right show the measured luminosities used to calculate SFRs. We show a (black dashed) line indicating SFR equivalence, as well as a (black dotted) line at the median IR to radio SFR excess of 2.5.}
\label{fig:sfr_comp}
\end{figure}

Figure \ref{fig:sfr_comp} shows a comparison between the SFRs estimated by our radio observations and those from the mid-IR WISE data for each of the galaxies in our sample. The radio-loud AGN has been excluded, but the remainder of galaxies do not exhibit such an extreme radio to IR excess as one would expect for a radio AGN. This figure also demonstrates that the SFR estimated from the IR is greater than that from the radio for a majority of the sample. We compute the distribution of the ratio of the IR to radio SFR using the Kaplan-Meier \citep[KM,][]{km} survival function implemented in \textit{ASURV} \citep{feig, iso} to account for non-detections and find that the IR SFR exceeds the radio SFR for  16 of 19 galaxies by a median factor of $\sim 2.5$. We investigate the origin of the IR to radio excess in the subsequent figures and in Section \ref{disc}.

\subsection{Trends With IR Colors}

\begin{figure}
\includegraphics[width=0.47\textwidth]{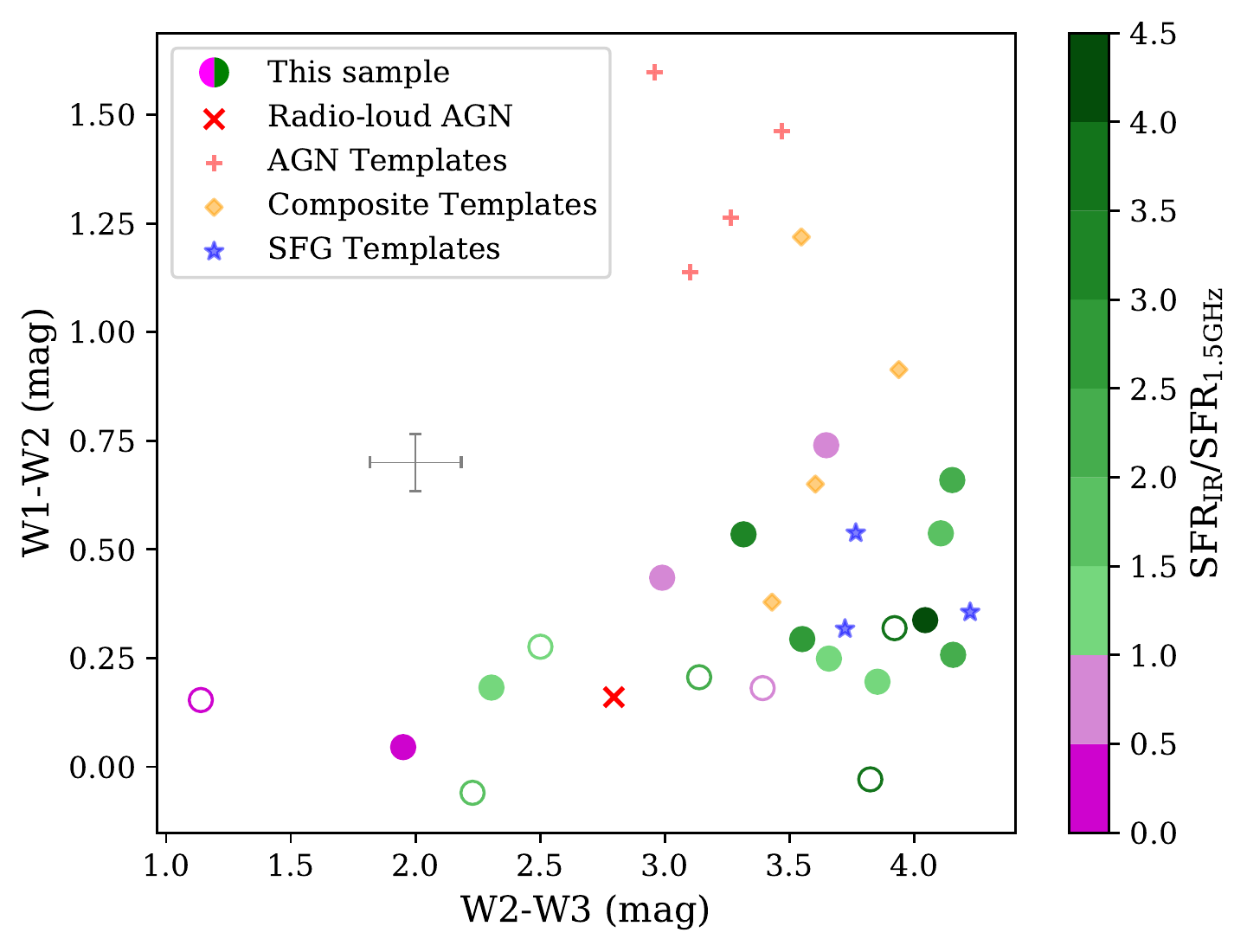}
\caption{The WISE W2--W3 versus W1--W2 colors for our sample. Points are colored according to the ratio of the IR SFR to the radio SFR, and hollow points indicate this ratio is a lower limit. A grey error bar gives the median uncertainty for the measured colors. Overlaid are model WISE colors of AGN, SFGs, and composite galaxies generated using IR templates from \citet{kirkpatrick}, redshifted to the median redshift of the sample ($z=0.6$). }
\label{fig:wisecolor}
\end{figure}

\begin{figure*}[t]
    \includegraphics[width=0.98\textwidth]{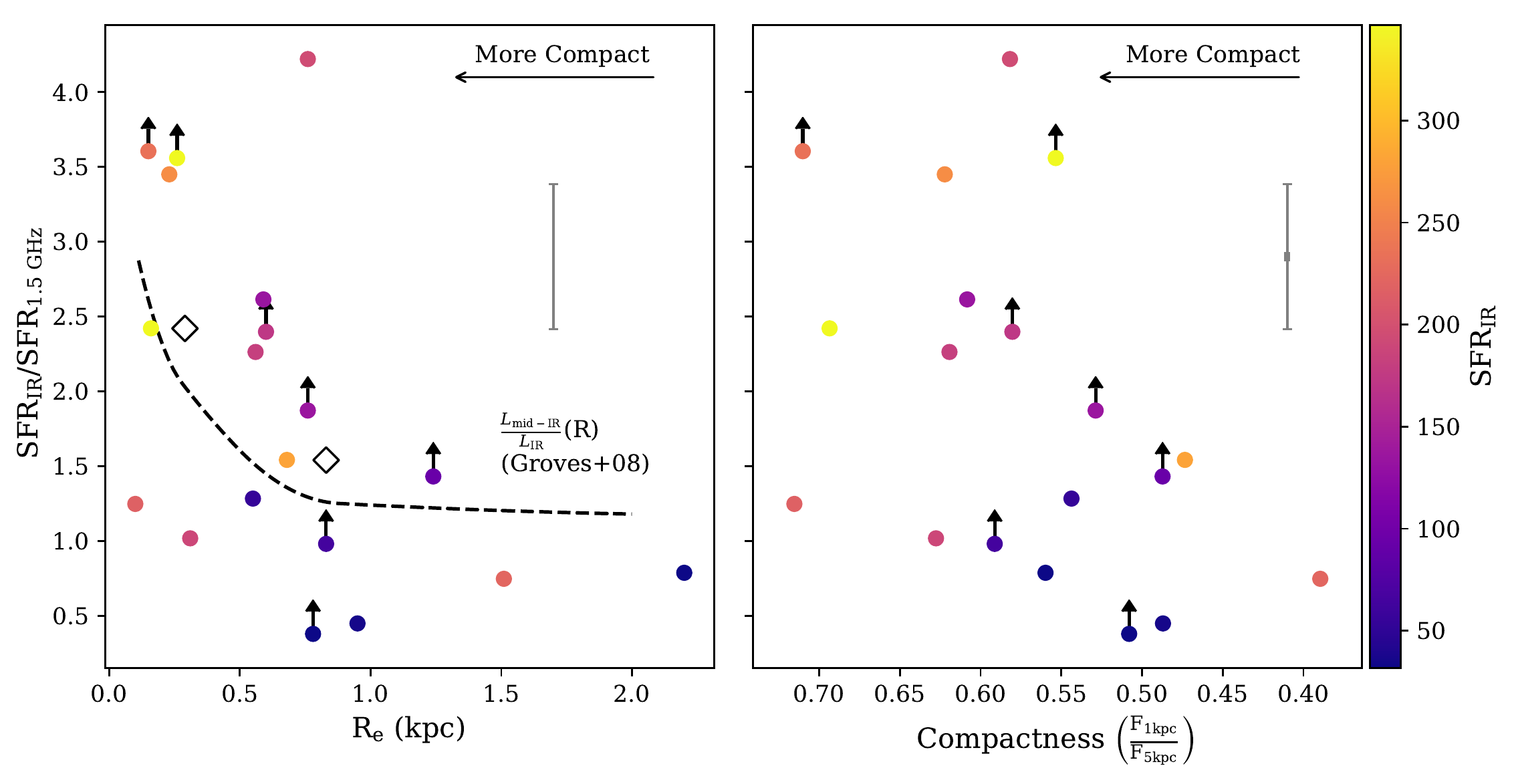}
    \caption{The ratio of the IR to the radio SFR versus the extent of the galaxy in the observed optical HST F814W filter, as measured in two ways. The left panel shows the ratio versus the effective radius derived from F814W/HST images. The right panel shows the same comparison, but measuring the extent using a compactness parameter (the ratio of optical flux within a radius of 1~kpc to within 5~kpc). The color bar indicates the IR SFR, which demonstrates that the observed trend is not wholly explained by the IR SFR. The markers with arrows indicate galaxies which were not detected in the radio, so the SFR ratio becomes a lower limit. The gray error bar shows the average uncertainty on the detected SFR ratio. The black diamonds show the median SFR ratio as computed by the KM estimator within two bins of effective radius. Both panels show that the most optically compact galaxies exhibit the highest IR to radio excess. We show a black dashed line indicating the scaling of the mid-IR to TIR with galaxy compactness as computed from \citet{groves}.} 
    \label{fig:radius}
\end{figure*}

Figure \ref{fig:wisecolor} is a W1--W2 versus W2--W3 color-color diagram for the WISE IR data for this sample. The position of a galaxy in the WISE color-color diagram can diagnose the mechanism dominating the IR emission (star formation vs AGN activity; e.g., \citealt{stern, assef, hickox}). Overlaid are model colors for AGN, SFG and composite templates from \citet{kirkpatrick}, redshifted to the median redshift ($z=0.6$) of the sample (varying the redshift across the range of our sample does not change the figure's result). This demonstrates that all of our sample falls within the SFG or composite regime in this color space, further illustrating a sub-dominant contribution from AGN activity to the mid-IR emission. This figure's color bar appears to show a weak trend that the galaxies with the highest IR to radio excesses are also the reddest at W2--W3. For context, we have added our radio-loud AGN (J0827) back into the sample for this figure. It does not exhibit AGN mid-IR colors, though this may be explained by the lack of UV-optical AGN signatures in the SDSS spectrum suggesting it is a radiatively inefficient AGN.

\subsection{Trends With Structural Parameters}

Figure \ref{fig:radius} compares the SFR derived from the IR to that derived from the radio, as a function of optical effective radius (left) and compactness, determined by the ratio of nuclear to total optical flux (right). It should be noted that because our sample galaxies occupy a small range in stellar mass, the effective radius serves as a proxy for compactness (Section \ref{sec:anc}). We compute the generalized Spearman rank correlation coefficient for the SFR ratio as a function of effective radius using \textit{ASURV} to account for lower limits, and find a coefficient of $\rho = -0.33$ with an associated p-value $p=0.16$. We thus only detect a correlation at the $\sim 1.4 \sigma$ level. We characterize this trend another way by binning the sample into two bins, those with effective radii equal to or less than the median radius and those with larger radii. We then compute the median SFR ratio in each bin using the KM estimator to account for lower limits and find that the median ratio is $\sim 1.6$ times larger for the galaxies with smaller effective radii. We thus conclude that the most compact galaxies exhibit a more extreme deviation from the IRRC. One physical interpretation of this is highlighted with a black dashed line in reference to \citet{groves}, which showed that more compact simulated galaxies have an enhanced brightness at mid-IR wavelengths for a fixed total IR luminosity due to hotter dust temperature distributions. The dashed line shown is a quadratic spline fit to the scaling of mid-IR (15~$\mu$m, the wavelength WISE W4 probes at $z\sim 0.5$) to IR luminosity with galaxy compactness as presented in \citet{groves}. This scaling is then fit for the normalization to our detected points. This scaling may help explain the observed IR to radio SFR excess, as a brightness enhancement at mid-IR wavelengths not modeled by our IR templates would cause an overestimation of the total IR luminosity and SFR for the most compact galaxies. This argument is somewhat tempered by the redshift dependence of the scaling, as our sample covers a redshift range $0.4 < z < 0.75$. Therefore we consider other physical interpretations of this plot, further discussed in Section \ref{disc}.

\begin{figure}[t]
    \includegraphics[width=0.47\textwidth]{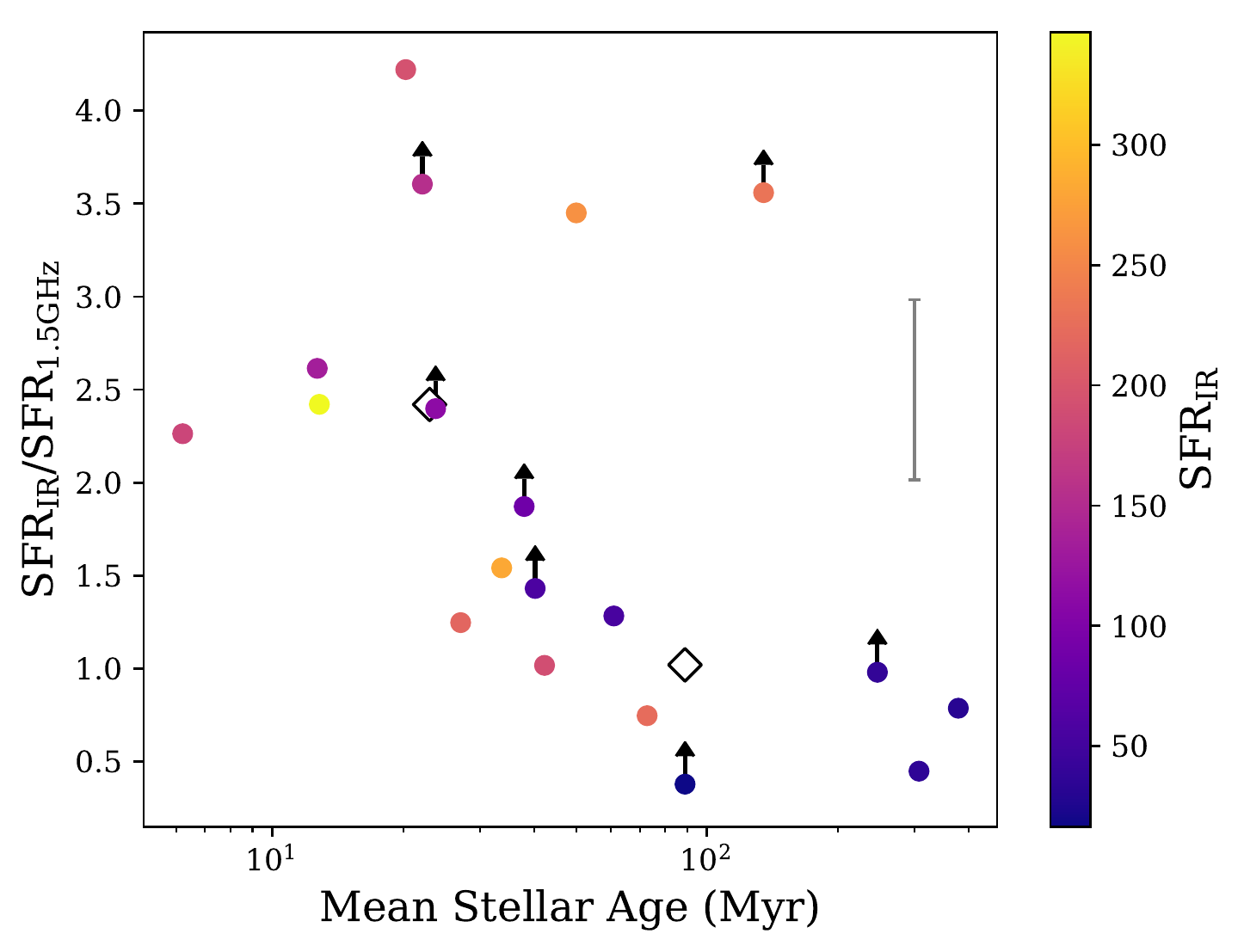}
    \caption{The IR to radio SFR ratio versus the mean stellar age. A gray vertical error bar shows the average uncertainty on the SFR ratio. The two black diamonds show the median SFR ratio computed by the KM estimator in two bins of mean stellar age. We observe a trend showing that the galaxies with the youngest populations of stars exhibit the greatest IR to radio excess. The color bar indicates that the trend is not explained by the IR SFR. }
    \label{fig:age}
\end{figure}

Figure \ref{fig:age} compares the IR to radio SFR ratio to the mean stellar age of the recent starburst. We again compute a Spearman rank coefficient using \textit{ASURV}, finding $\rho = -0.46$ with an associated p-value $p=0.05$. Thus, we detect a $\sim 2\sigma$ inverse correlation of the SFR ratio with mean stellar age. We also bin the sample by the median mean stellar age and compute the median SFR ratio for each bin using the KM estimator, finding that galaxies with younger stellar populations exhibit SFR ratios $\sim 2.4$ times greater than galaxies with older stellar populations. This figure, along with Figure \ref{fig:radius}, suggests that young compact starburst galaxies either exhibit more IR or less radio luminosity than their instantaneous SFRs would imply in regular SFGs.

\begin{figure}[t]
    \centering
    \includegraphics[width=0.47\textwidth]{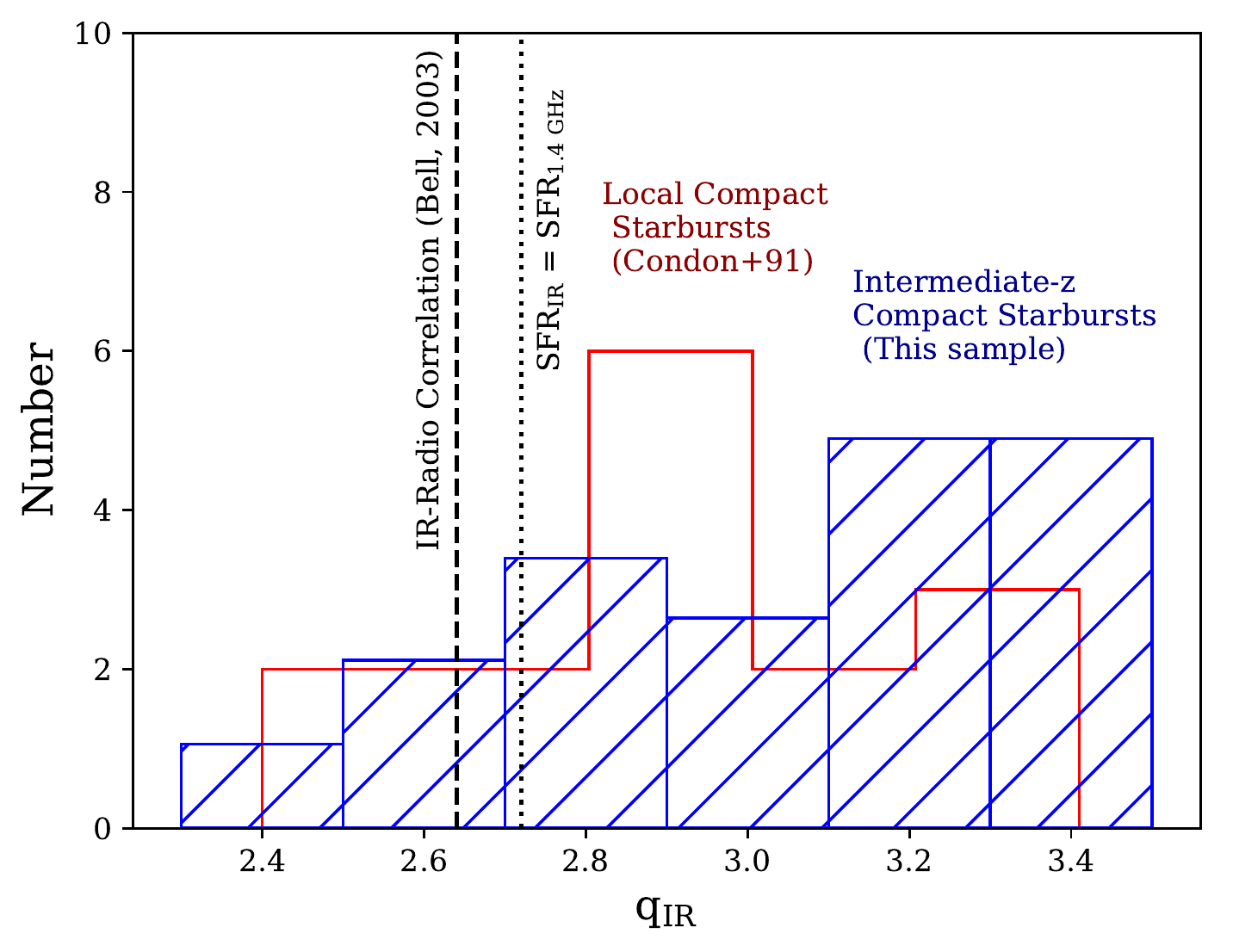}
    \caption{The histogram of $q_{\mathrm{IR}}$ values for both this sample and the sample of local compact starbursts studied in \citet{lirg}. The $q_{\mathrm{IR}}$ distribution of our sample has been generated by the KM estimator to account for lower limits. We show the $q_{\mathrm{IR}}$ value corresponding to the IRRC as measured by \citet[][black dashed line]{bell} as well as the $q_{\mathrm{IR}}$ value implied by equating equations \ref{eq:sfr} and \ref{eq:IRSFR} (black dotted line). Our sample galaxies are similarly deviant from the IRRC, with a tail towards even more extreme ratios. We reference this figure in Section \ref{disc} to argue that radio continuum emission in our sample may be suppressed by free-free absorption just as in local compact starbursts.}
    \label{fig:condon}
\end{figure}

In Figure \ref{fig:condon} we compare the histograms of the $q_{\mathrm{IR}}$ parameter for our sample:

\begin{equation}
    q_{\mathrm{IR}} = \log_{10}\left(\frac{L_{\mathrm{IR}}}{3.75\times 10^{12} \ \mathrm{W}}\right) - \log_{10}\left(\frac{L_{1.4\mathrm{GHz}}}{\mathrm{W \ Hz}^{-1}}\right),
\end{equation}
defined as the logarithmic ratio of the IR to 1.4 GHz luminosity, to the $q_{\mathrm{IR}}$ for a sample of local compact starbursts studied in \citet{lirg}. The $q_{\mathrm{IR}}$ parameter was originally defined using the far-IR luminosity ($\sim 40$--$120 \ \mu$m), which is the convention used in \citet{lirg}. \citet{Yun} found the average value of this parameter to be $q_{\mathrm{FIR}} = 2.34$ in local galaxies. Using the integrated IR luminosity however yields a value of $q_{\mathrm{IR}} = 2.64$ \citep{bell}. In order to compare our sample to local compact starbursts, we therefore add 0.3 dex to the $q_{\mathrm{FIR}}$ values presented in \citet{lirg} to convert them to the $q_{\mathrm{IR}}$ convention. We also use the chosen non-thermal spectral index of $\alpha^{NT} = -0.8$ to extrapolate our 1.5~GHz luminosities to 1.4~GHz. Finally, we use \textit{ASURV}'s implementation of the differential KM estimator to output the expected distribution of $q_{\mathrm{IR}}$ given lower limits. We find that our sources are similarly deviant from the IRRC as local compact starbursts, with a tail towards even \textit{more} extreme ratios. In Section \ref{disc}, we use this result to argue that radio continuum attenuation by free-free absorption is one of the favored explanations for the observed SFR discrepancy. This is because \citet{lirg} found that local compact starbursts are optically thick at GHz frequencies due to free-free absorption in H II regions. Only after correcting for free-free absorption did their sources obey the IRRC.

\subsection{Trends With Outflow Characteristics}

\begin{figure*}[t]
\includegraphics[width=0.97\textwidth]{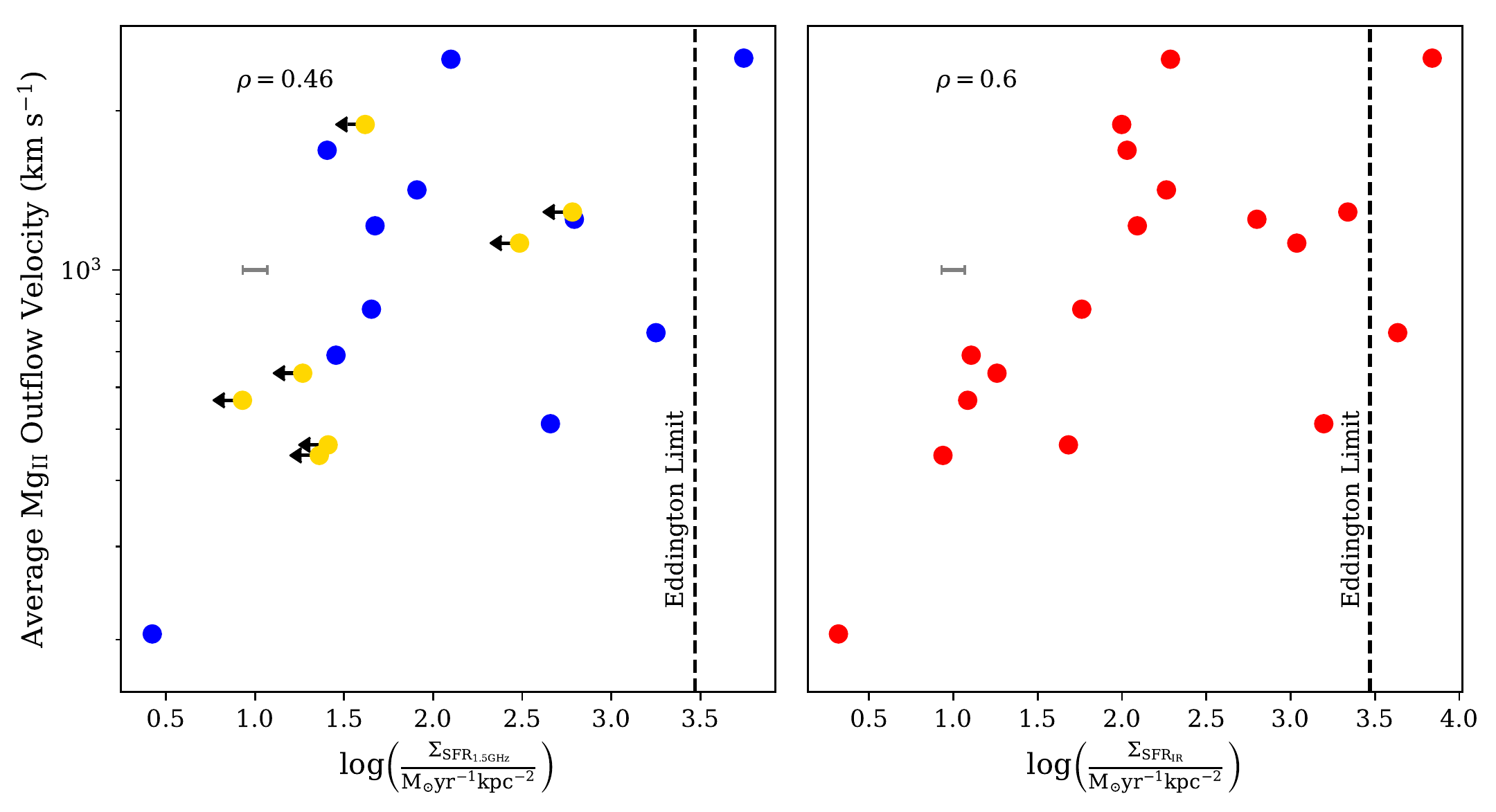}
\caption{The average outflow speed as traced by the Mg II absorption versus star formation surface density as traced by the radio (left panel) and IR (right panel). The sizes used for the surface density are effective radii measured by HST. A grey horizontal error bar shows the average uncertainty in surface density. There is a positive correlation between star formation density and outflow speed, which is stronger for the SFRs determined by the IR emission (generalized Spearman rank $\rho$ values). The galaxy without a Mg II detection is excluded, but this source is known to host an extreme outflow (see text).}
\label{fig:speed}
\end{figure*}

To investigate whether compact starburst activity is connected to the observed outflows, we compare SFR surface density as traced by the radio and IR with outflow velocity as traced by Mg II absorption lines in Figure \ref{fig:speed}. The source (J2118) with no detected Mg II line (possibly due to emission filling) is excluded from the plot, but this does not imply it lacks outflow activity. This source - the galaxy with the largest average SFR - in fact hosts an extreme (50~kpc radius) ionized outflow observed through spatially resolved emission line measurements from integral field unit observations \citep{rupke}. We find a positive correlation between SFR surface density and outflow velocity, suggestive of star formation feedback. We compute the generalized Spearman rank coefficient with \textit{ASURV} for each SFR indicator and find that the correlation is stronger for SFRs determined by the IR ($\rho = 0.6$, $p=0.008$) compared to those from the radio ($\rho = 0.46$, $p=0.06$). We thus detect a correlation between SFR surface density and outflow velocity at a higher significance from the IR ($\sim 2.7\sigma$) compared to the radio ($\sim 1.9\sigma$). If these galaxies are undergoing star formation feedback, we would expect a correlation between SFR surface density and gaseous outflow speed. Indeed, these quantities are known to be correlated (see \citet{rubin} for a study of galaxies at comparable redshift with much smaller outflow velocities and $\Sigma_{\mathrm{SFR}}$, or \citet{heckman} for a compliation of our sample with lower surface density galaxies). This tentatively suggests that the IR is tracing the instantaneous SFR more accurately than the radio emission, but FIR data as a third SFR calibrator as well as larger sample sizes are required to test this conclusively.

\section{Discussion\label{disc}}
In Section \ref{sec:results}, we showed that 16 out of 19 of our sources exhibit enhanced IR SFRs compared to radio estimates, and that this deviation is maximized for the most compact galaxies and the galaxies with the smallest mean stellar ages. This suggests that compact starbursts deviate from the IRRC. In this context, we identify four physical effects that could produce the discrepancy between the radio and IR SFRs:

\textit{Free-free absorption}: In this picture, radio continuum emission is reabsorbed by free-free processes in H II regions. Evidence for this picture is shown in Figure \ref{fig:condon}, which illustrates that our sample galaxies exhibit even more extreme deviations from the IRRC compared to compact starbursts in the local universe. \citet{lirg} showed that local compact starbursts are optically thick at GHz frequencies due to free-free absorption, and that the sample only obeyed the IRRC once the radio flux had been corrected for this absorption. We are currently unable to make these corrections, as we lack the spectral index data to do so. Figure \ref{fig:radius} shows that the galaxies with the highest IR to radio excess are the most compact, which in turn would produce a higher column density of absorbing electrons, consistent with this scenario.

Synchrotron self-absorption should not be important in our sample galaxies due to their low brightness temperatures. As our sources are unresolved, we can only estimate a lower limit on the brightness temperature. However, these lower limits are very small, with a median of $T_{b} = 10$ K and a maximum of $T_{b}=68$ K. According to \citet{essential}, a self-absorbed synchrotron source at 1.5~GHz with a brightness temperature of $T_{b} = 10$ K would require a galactic magnetic field strength of $B \sim 10^{18}$ G, which is unfeasible even as an upper limit. We thus do not consider synchrotron self-absorption as a possible explanation of the deviation from the IRRC.

\textit{Hot dust scenario}: The correlation between the deviation from the IRRC and compactness may also be consistent with changes in dust temperature. Thus, the observed discrepancy may stem from a mid-IR enhancement not accounted for by our IR templates, as simulated compact galaxies exhibit a brightness enhancement at mid-IR wavelengths for a fixed total IR luminosity \citep{groves}. This is because dust temperature distributions depend on the specific photon density, so compact star-forming regions produce hotter dust grain distributions, shifting the IR energy output to shorter wavelengths. The dashed line shown in Figure \ref{fig:radius} indicates this scaling of mid-IR enhancement with compactness, and seems to fit our points well. This argument is tempered however by the fact that this effect is highly redshift dependent.


\textit{Tracer timescales scenario}: The discrepancy may also stem from the fact that the two SFR indicators trace star formation on different timescales. To begin producing synchrotron emission, a nascent starburst needs enough time for the massive stars to explode in supernovae and accelerate cosmic ray electrons, whereas the IR traces star formation nearly instantaneously. Neglecting other considerations such as galactic magnetic field strength, one would expect an IR to radio excess for at least $\sim 4$ Myr after the birth of a starburst \citep{roussel}. Because this sample preferentially hosts very young stellar populations, we may be observing a portion of the sample within this few Myr window of IR excess, which may be supported by Figure \ref{fig:age}. 

Another timescale discrepancy argument stems from the result that IR emission from simulated mergers begins to overestimate the SFR during epochs of rapid decline in the SFR, such as quenching events \citep{hayward}. This is because IR emission as an SFR indicator traces star-formation averaged over $\sim 100$ Myr timescales. As the SFR begins to decline during a quenching event, the IR emitting dust remains hot, such that the observed IR luminosity begins to progressively overestimate the instantaneous SFR. Meanwhile, synchrotron emission traces star-formation on $\sim 10$ Myr timescales. If we are indeed observing these galaxies during a quenching event triggered by outflows, then the observed IR emission may be overestimating the true SFRs.

\textit{Convective wind scenario}: An alternative scenario that could produce the observed deviation is the presence of convective winds in our sample galaxies. \citet{lisenfeld} showed that a local dwarf post-starburst galaxy's radio spectrum is consistent with the expulsion of cosmic ray electrons from the galaxy via a convective wind. Such a removal of cosmic rays could suppress the synchrotron emission associated with star formation. Without radio observations at multiple frequencies, we cannot currently rule out this possibility.

In the free-free absorption scenario, the first tracer-timescale scenario, or in the case of a convective wind, the observed radio flux would underestimate the true SFR. Alternatively, the IR would overestimate the instantaneous SFR in the case of hot dust, or in the second tracer-timescale scenario. The SFR discrepancy we observe in our sample may result from some combination of these physical pictures. The stronger connection between IR luminosity and outflow velocity (Figure \ref{fig:speed}) may give credence to the IR being the better tracer of SF. However, a robust assessment of this will require future FIR observations (with SOFIA) to probe emission near the dust peak. These observations would constrain the appropriate IR SED templates and thus SFRs of these galaxies significantly.

If these galaxies are indeed optically thick to free-free emission, IR emission may be suppressed as well. This extinction would be more substantial at shorter wavelengths, causing a reddening of the IR SED. If our IR templates fail to account for this reddening, we may be underestimating not only the radio SFR but also the IR SFR. In this case, the true SFRs would imply the sample lies even closer to the Eddington star formation limit than our current estimates suggest.


\section{Conclusions}

This paper presents star formation rates calculated from 1.5~GHz (rest-frame $\sim$2.4~GHz) continuum observations of a sample of massive ($M_{*} \sim 10^{11} \ \mathrm{M_{\odot}}$), compact starburst galaxies (at $z\sim 0.6$) hosting high-velocity gaseous outflows ($|v| \gtrsim 10^{3}$~km~s$^{-1}$) thought to be triggered by star formation activity. These observations provide an independent measurement of these galaxies' SFRs, as radio emission is optically thin to dust absorption, and calculation of SFRs in the non-thermal regime relies on different assumptions.

We found that only one object exhibited characteristics of an AGN at radio frequencies, further constraining AGN activity in the sample. We also found that the SFR estimates from the IR exceeded those from the radio for 16 out of 19 galaxies by a median factor of 2.5. The galaxies which exhibit the highest IR to radio excess are the most compact in optical morphology and host the youngest populations of stars, suggesting that compact starbursts deviate from the IRRC.

We conclude that the origin of this deviation could be due to several possible physical effects: free-free absorption in H II regions stifling synchrotron emission produced by star formation, an excess of IR luminosity due to hot dust, a difference in the timescales over which each SFR indicator traces star formation, a convective wind, or some combination of the above.

Upcoming FIR observations to constrain the luminosity near the dust peak (currently limited to the two brightest sources in the sample) should provide a third calibrator for these galaxies' SFRs. With accurate estimates of these SFRs, we may be able to probe star formation feedback at its most extreme, as well as use the sample as an analog to understand massive galaxy evolution at high redshift.


\acknowledgments
GCP acknowledges support from the Dartmouth Fellowship. AAK would like to thank Drew Medlin for his assistance wrangling the initial pipeline data products and James Condon for helpful conversations. This material is based upon work supported by the National Science Foundation (NSF) under a collaborative grant (AST-1814233, 1813299, 1813365,
1814159 and 1813702). GCP was a summer student at the National Radio Astronomy Observatory. The National Radio Astronomy Observatory is a facility of the National Science Foundation operated under cooperative agreement by Associated Universities, Inc.

%

\vspace{5mm}
\facilities{VLA(NRAO), HST(STIS)}


\software{\citep[astropy, ][]{astropy}, \ \citep[CASA v. 5.3.0; v. 5.4.1-32, ][]{Casa}}



\begin{figure}[h]
\begin{centering}
\includegraphics[width=.95\textwidth]{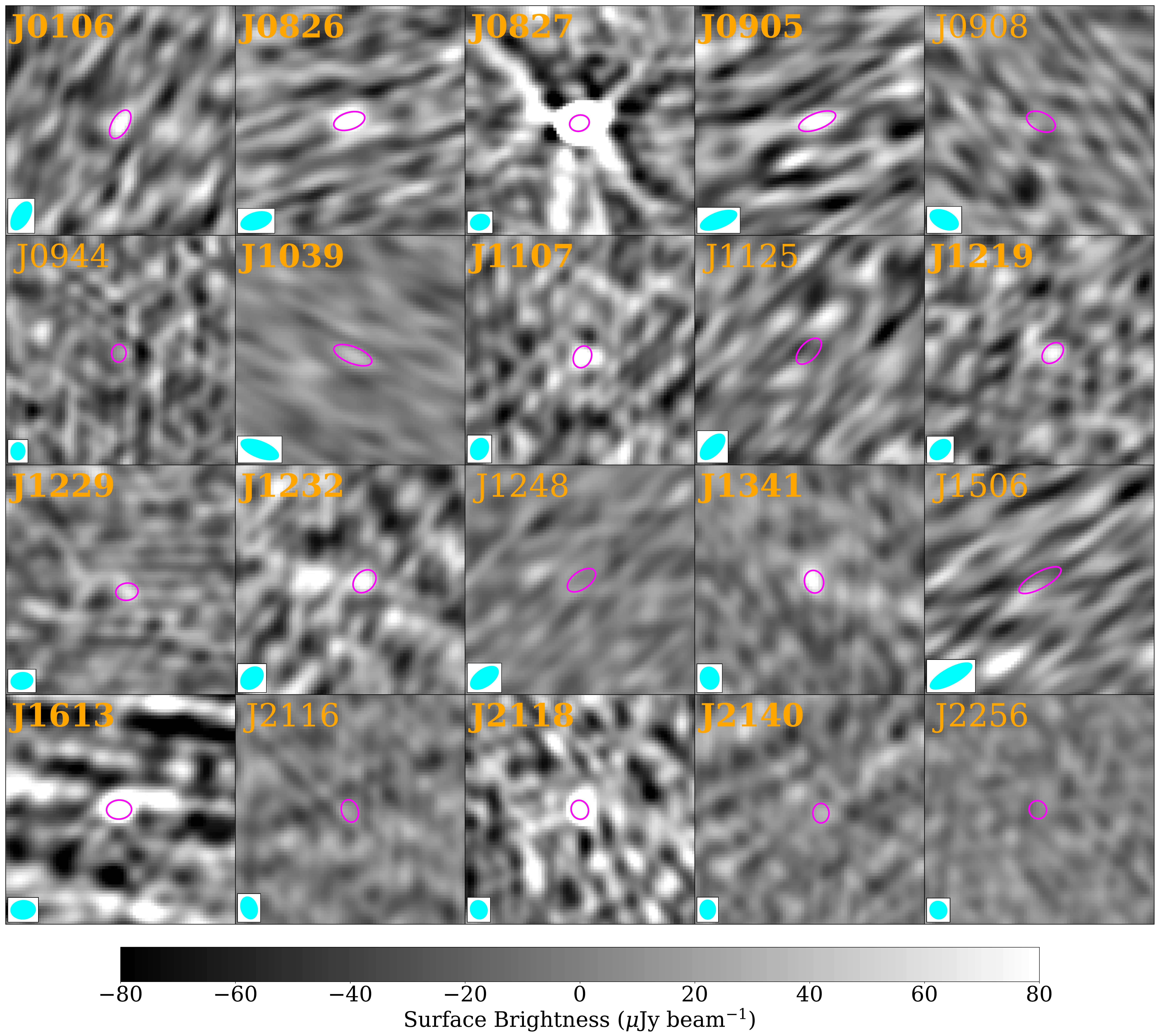}
\caption{$15 \  \sq  \arcsec$ cutouts of cleaned VLA images with which fluxes are measured. The images are centered according to the corresponding HST centroid.  Names in bold are deemed detections by the algorithm discussed in Section \ref{flux-meas}. The synthesized beam size is shown in the lower-left corner in cyan and the Gaussian 1$\sigma$ source fit width is highlighted with magenta ellipses. If there exists a pixel brighter than 3 times the image noise within 1$\arcsec$ of the image center, that position is fixed to be the center of the Gaussian fit. Otherwise, the Gaussian is fixed at the HST centroid position. }
\end{centering}
\label{fig:Stamps}
\end{figure}

\begin{figure}[h]
\begin{centering}
\includegraphics[width=.95\textwidth]{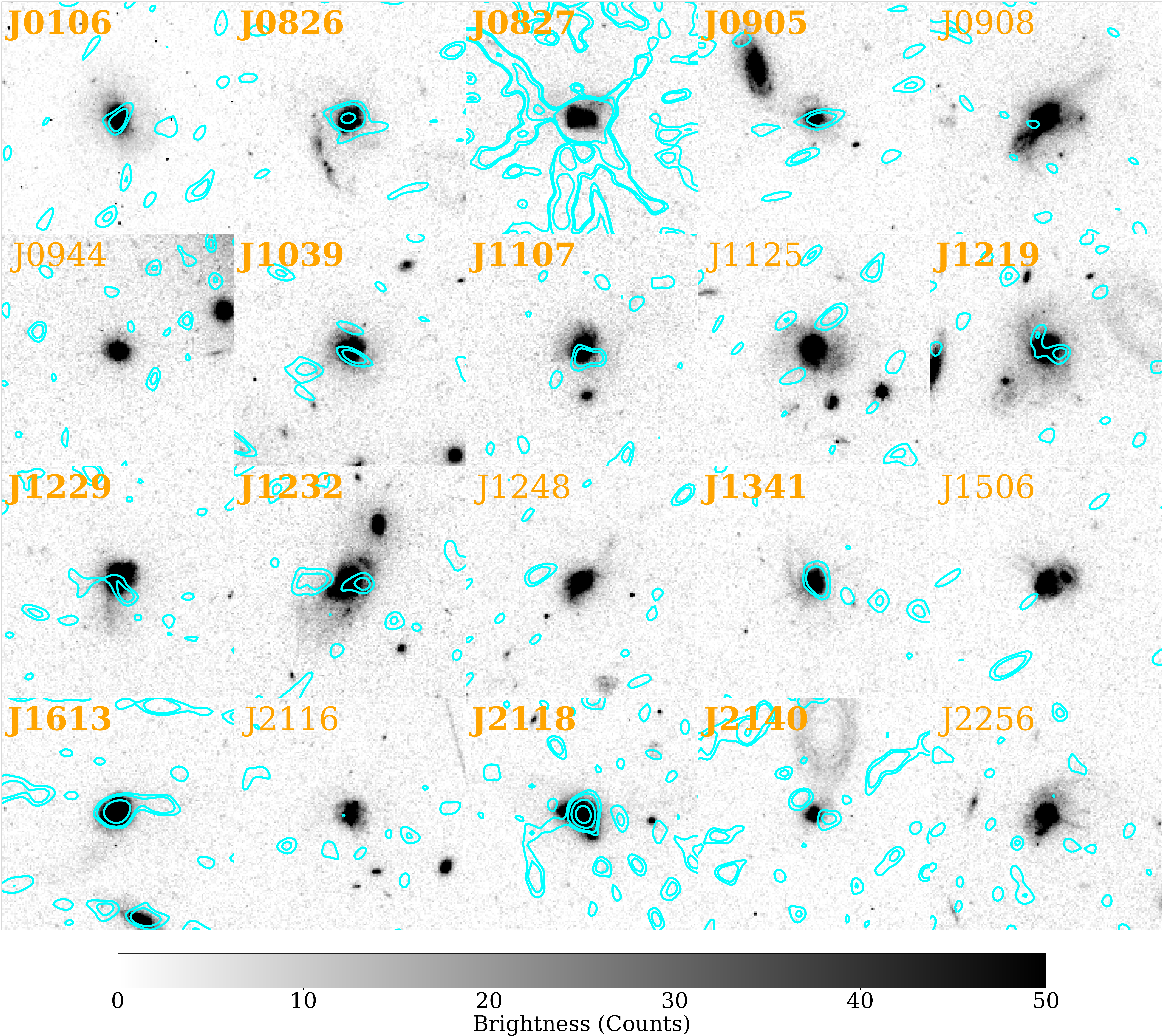}
\caption{HST F814W images of our sample galaxies with contours drawn by the radio image. Contours are drawn on a logarithmic scale, including 2, 3, 6, and 12$\sigma$ contours. Here, we demonstrate that there is significant radio emission overlapping the optical bounds of the galaxy for the galaxies our algorithm deemed detections (bolded names).}
\end{centering}
\label{fig:HST_Stamps}
\end{figure}

\appendix
Here we outline details of the radio observations and present derived galaxy quantities. Table \ref{table:da} displays calculated quantities for the sample galaxies. Figure \ref{fig:Stamps} shows cutouts of our images synthesized from VLA data with Gaussian source fits overlaid, and Figure \ref{fig:HST_Stamps} shows HST images of the sample, with the corresponding radio image contours drawn over them.

\tabcolsep=0.26cm
\small
\begin{deluxetable}{cccccccccc}[h]
\caption{The galaxies in the radio sample and measured quantities including the redshift, 1.5~GHz flux, luminosity \& SFR, the 12 \& 22 $\mu$m fluxes, the total IR luminosity and the IR SFR.}
\label{table:da}

\tablehead{\colhead{Name} & \colhead{z} & \colhead{$F_{1.5\mathrm{GHz}}$} & \colhead{$L_{1.5\mathrm{GHz}}$} & \colhead{SFR$_{1.5\mathrm{GHz}}$} & \colhead{$F_{12 \mu \mathrm{m}}$} & \colhead{$F_{22 \mu \mathrm{m}}$} & \colhead{$L_{\mathrm{IR}}$} & \colhead{SFR$_{\mathrm{IR}}$}\\ \colhead{ } & \colhead{ } & \colhead{$\mu \mathrm{Jy}$} & \colhead{$10^{22} \ \mathrm{W} \ \mathrm{Hz}^{-1}$} & \colhead{$\mathrm{M_{\odot}\,yr^{-1}}$} & \colhead{$\mathrm{mJy}$} & \colhead{$\mathrm{mJy}$} & \colhead{$10^{36}$ W} & \colhead{$\mathrm{M_{\odot}\,yr^{-1}}$ }}
\startdata
J0106--1023 & 0.454 & $89\pm19$ & $6.3 \pm 1.4$ & $52 \pm 11$ & $1.28 \pm 0.16$ & $2.78 \pm 0.94$ & $348 \pm 54$ & $135 \pm 21$ \\
J0826+4305 & 0.603 & $168\pm20$ & $23.1 \pm 2.8$ & $187 \pm 23$ & $0.80 \pm 0.12$ & $3.23 \pm 0.81$ & $491 \pm 75$ & $191 \pm 29$ \\
J0827+2954 & 0.681 & $16500\pm12$ & $3017.3 \pm 2.1$ & --- & $0.24 \pm 0.12$ & $1.24 \pm 0.81$ & $207 \pm 31$ & $80 \pm 12$ \\
J0905+5759 & 0.711 & $107\pm22$ & $21.6 \pm 4.4$ & $174 \pm 36$ & $0.62 \pm 0.10$ & $2.40 \pm 0.67$ & $560 \pm 83$ & $217 \pm 32$ \\
J0908+1039 & 0.502 & $\leq 56$ & $\leq 5.0$ & $\leq 41$ & $0.45 \pm 0.11$ & $0.25 \pm 0.75$ & $152 \pm 23$ & $58.8 \pm 8.8$ \\
J0944+0930 & 0.514 & $\leq 56$ & $\leq 5.2$ & $\leq 43$ & $0.99 \pm 0.13$ & $3.37 \pm 0.89$ & $398 \pm 62$ & $154 \pm 24$ \\
J1039+4537 & 0.634 & $34 \pm 10$ & $5.3 \pm 1.5$ & $43 \pm 12$ & $0.30 \pm 0.10$ & --- & $142 \pm 25$ & $54.9 \pm 9.7$ \\
J1107+0417 & 0.467 & $130 \pm 25$ & $9.8 \pm 1.9$ & $80 \pm 15$ & $1.55 \pm 0.12$ & $4.17 \pm 0.79$ & $466 \pm 74$ & $181 \pm 29$ \\
J1125--0145 & 0.519 & $\leq 60$ & $\leq 5.8$ & $\leq 47$ & $0.72 \pm 0.12$ & $1.98 \pm 0.79$ & $290 \pm 45$ & $113 \pm 17$ \\
J1219+0336 & 0.451 & $81 \pm 22$ & $5.6 \pm 1.5$ & $46 \pm 13$ & $1.76 \pm 0.13$ & $6.29 \pm 0.79$ & $501 \pm 81$ & $195 \pm 32$ \\
J1229+3545 & 0.614 & $70 \pm 16$ & $10.0 \pm 2.3$ & $81 \pm 19$ & $0.17 \pm 0.11$ & $0.10 \pm 0.78$ & $93 \pm 14$ & $36.2 \pm 5.3$ \\
J1232+0723 & 0.401 & $93 \pm 25$ & $4.9 \pm 1.3$ & $40 \pm 11$ & $0.62 \pm 0.15$ & --- & $82 \pm 18$ & $31.7 \pm 7.0$ \\
J1248+0601 & 0.632 & $\leq 35$ & $\leq 5.4$ & $\leq 44$ & $0.05 \pm 0.13$ & $0.44 \pm 0.84$ & $42.7 \pm 6.8$ & $16.6 \pm 2.7$ \\
J1341--0321 & 0.661 & $104 \pm 16$ & $17.7 \pm 2.7$ & $143 \pm 22$ & $1.184 \pm 0.094$ & $4.58 \pm 0.63$ & $890\pm 130$ & $347 \pm 52$ \\
J1506+6131 & 0.437 & $\leq 76$ & $\leq 4.9$ & $\leq 40$ & $0.442 \pm 0.059$ & $0.34 \pm 0.42$ & $101 \pm 16$ & $39.3 \pm 6.2$ \\
J1613+2834 & 0.449 & $324 \pm 26$ & $22.3 \pm 1.8$ & $183 \pm 15$ & $2.612 \pm 0.082$ & $10.06 \pm 0.66$ & $730 \pm 120$ & $282 \pm 48$ \\
J2116--0634 & 0.728 & $\leq 38$ & $\leq 8.0$ & $\leq 65$ & $0.60 \pm 0.15$ & --- & $595 \pm 90$ & $231 \pm 35$ \\
J2118+0017 & 0.459 & $505 \pm 25$ & $36.6 \pm 1.8$ & $300 \pm 15$ & $2.02 \pm 0.13$ & $5.15 \pm 0.85$ & $576 \pm 91$ & $224 \pm 35$ \\
J2140+1209 & 0.751 & $41 \pm 10$ & $9.4 \pm 2.2$ & $76 \pm 18$ & $0.67 \pm 0.12$ & $2.05 \pm 0.86$ & $673 \pm 99$ & $261 \pm 38$ \\
J2256+1504 & 0.727 & $\leq 27$ & $\leq 5.8$ & $\leq 47$ & $0.21 \pm 0.12$ & $1.57 \pm 0.79$ & $225 \pm 34$ & $87 \pm 13$
\enddata
\end{deluxetable}

\FloatBarrier




\begin{thebibliography}{}
\small{

\bibitem[Assef et al.(2013)]{assef} Assef, R.~J., Stern, D., Kochanek, C.~S., et al.\ 2013, \apj, 772, 26

\bibitem[Bell(2003)]{bell} Bell, E.~F.\ 2003, \apj, 586, 794


\bibitem[Benson et al.(2003)]{Benson} Benson, A.~J., Bower, R.~G., Frenk, C.~S., et al.\ 2003, \apj, 599, 38.





\bibitem[Condon et al.(1991)]{lirg} Condon, J.~J., Huang, Z.-P., Yin, Q.~F., et al.\ 1991, \apj, 378, 65

\bibitem[Condon(1992)]{radiorev} Condon, J.~J.\ 1992, \araa, 30, 575


\bibitem[Condon(1997)]{fits} Condon, J.~J.\ 1997, \pasp, 109, 166


\bibitem[Condon et al.(1998)]{Vlass} Condon, J.~J., Cotton, W.~D., Greisen, E.~W., et al.\ 1998, \aj, 115, 1693.
\bibitem[Condon \& Matthews(2018)]{condon} Condon, J.~J., \& Matthews, A.~M.\ 2018, \pasp, 130, 073001
\bibitem[Condon \& Ransom(2016)]{essential} Condon, J.~J. \& Ransom, S.~M.\ 2016, Essential Radio Astronomy, by James J. Condon and Scott M. Ransom. ISBN: 978-0-691-13779-7. Princeton, NJ: Princeton University Press, 2016.

\bibitem[de Jong et al.(1985)]{Jong} de Jong, T., Klein, U., Wielebinski, R., et al.\ 1985, \aap, 147, L6.
\bibitem[Croton et al.(2006)]{Croton} Croton, D.~J., Springel, V., White, S.~D.~M., et al.\ 2006, \mnras, 365, 11.
\bibitem[Diamond-Stanic et al.(2012)]{Diamond2012} Diamond-Stanic, A.~M., Moustakas, J., Tremonti, C.~A., et al.\ 2012, \apj, 755, L26.

\bibitem[Feigelson \& Nelson(1985)]{feig} Feigelson, E.~D., \& Nelson, P.~I.\ 1985, \apj, 293, 192


\bibitem[Ferrarese \& Merritt(2000)]{Ferr} Ferrarese, L. \& Merritt, D.\ 2000, \apj, 539, L9.

\bibitem[Gebhardt et al.(2000)]{geb} Gebhardt, K., Bender, R., Bower, G., et al.\ 2000, \apjl, 539, L13


\bibitem[Geach et al.(2014)]{geach1} Geach, J.~E., Hickox, R.~C., Diamond-Stanic, A.~M., et al.\ 2014, Nature, 516, 68

\bibitem[Geach et al.(2018)]{geach2} Geach, J.~E., Tremonti, C., Diamond-Stanic, A.~M., et al.\ 2018, The Astrophysical Journal, 864, L1

\bibitem[Groves et al.(2008)]{groves} Groves, B., Dopita, M.~A., Sutherland, R.~S., et al.\ 2008, \apjs, 176, 438


\bibitem[Hayward et al.(2014)]{hayward} Hayward, C.~C., Lanz, L., Ashby, M.~L.~N., et al.\ 2014, Monthly Notices of the Royal Astronomical Society, 445, 1598

\bibitem[Heckman \& Borthakur(2016)]{heckman} Heckman, T.~M., \& Borthakur, S.\ 2016, \apj, 822, 9


\bibitem[Helou et al.(1985)]{Helou} Helou, G., Soifer, B.~T. \& Rowan-Robinson, M.\ 1985, \apj, 298, L7.

\bibitem[Helou \& Bicay(1993)]{hebic} Helou, G. \& Bicay, M.~D.\ 1993, \apj, 415, 93


\bibitem[Hickox et al.(2017)]{hickox} Hickox, R.~C., Myers, A.~D., Greene, J.~E., et al.\ 2017, \apj, 849, 53



\bibitem[Isobe et al.(1986)]{iso} Isobe, T., Feigelson, E.~D., \& Nelson, P.~I.\ 1986, \apj, 306, 490

\bibitem[Kaplan \& Meier(1958)]{km} Kaplan, E.~L., \& Meier, P.\ 1958, J. Am. Statistical Association, 53, 457.


\bibitem[Kepley et al.(2020)]{kepley} Kepley, A.~A., Tsutsumi, T., Brogan, C.~L., et al.\ 2020, \pasp, 132, 024505



\bibitem[Kirkpatrick et al.(2015)]{kirkpatrick} Kirkpatrick, A., Pope, A., Sajina, A., et al.\ 2015, The Astrophysical Journal, 814, 9

\bibitem[Kormendy \& Ho(2013)]{kor} Kormendy, J., \& Ho, L.~C.\ 2013, \araa, 51, 511



\bibitem[Kroupa(2001)]{Kroupa} Kroupa, P.\ 2001, \mnras, 322, 231 
\bibitem[Lacki et al.(2010)]{lacki} Lacki, B.~C., Thompson, T.~A., \& Quataert, E.\ 2010, \apj, 717, 1




\bibitem[Lehnert, \& Heckman(1996)]{lehnert} Lehnert, M.~D., \& Heckman, T.~M.\ 1996, The Astrophysical Journal, 472, 546

\bibitem[Lisenfeld et al.(2004)]{lisenfeld} Lisenfeld, U., Wilding, T.~W., Pooley, G.~G., et al.\ 2004, \mnras, 349, 1335

\bibitem[Lutz et al.(2016)]{lutz} Lutz, D., Berta, S., Contursi, A., et al.\ 2016, \aap, 591, A136



\bibitem[McMullin et al.(2007)]{Casa} McMullin, J.~P., Waters, B., Schiebel, D., Young, W., \& Golap, K.\ 2007, Astronomical Data Analysis Software and Systems XVI, 376, 127 




\bibitem[Meurer et al.(1997)]{meurer} Meurer, G.~R., Heckman, T.~M., Lehnert, M.~D., et al.\ 1997, The Astronomical Journal, 114, 54



\bibitem[Murphy(2011)]{murphy11} Murphy, E. J., Condon, J. J., Schinnerer, E., et al. 2011, ApJ, 737, 67
\bibitem[Murray et al.(2005)]{murray} Murray, N., Quataert, E., \& Thompson, T.~A.\ 2005, The Astrophysical Journal, 618, 569

\bibitem[Roussel et al.(2003)]{roussel} Roussel, H., Helou, G., Beck, R., et al.\ 2003, \apj, 593, 733

\bibitem[Rubin et al.(2014)]{rubin} Rubin, K.~H.~R., Prochaska, J.~X., Koo, D.~C., et al.\ 2014, \apj, 794, 156


\bibitem[Rupke et al.(2019)]{rupke} Rupke, D.~S.~N., Coil, A., Geach, J.~E., et al.\ 2019, arXiv e-prints, arXiv:1910.13507


\bibitem[Schlafly et al.(2019)]{unwise} Schlafly, E.~F., Meisner, A.~M., \& Green, G.~M.\ 2019, \apjs, 240, 30


\bibitem[Sell et al.(2014)]{sell} Sell, P.~H., Tremonti, C.~A., Hickox, R.~C., et al.\ 2014, Monthly Notices of the Royal Astronomical Society, 441, 3417

\bibitem[Smol{\v{c}}i{\'c} et al.(2017)]{smolcic} Smol{\v{c}}i{\'c}, V., Novak, M., Delvecchio, I., et al.\ 2017, \aap, 602, A6


\bibitem[Stern et al.(2012)]{stern} Stern, D., Assef, R.~J., Benford, D.~J., et al.\ 2012, \apj, 753, 30

\bibitem[Tabatabaei et al.(2017)]{taba} Tabatabaei, F.~S., Schinnerer, E., Krause, M., et al.\ 2017, \apj, 836, 185


\bibitem[The Astropy Collaboration et al.(2018)]{astropy} The Astropy Collaboration, Price-Whelan, A.~M., Sip{\H o}cz, B.~M., et al.\ 2018, arXiv:1801.02634 
\bibitem[Thompson et al.(2005)]{thompson} Thompson, T.~A., Quataert, E., \& Murray, N.\ 2005, The Astrophysical Journal, 630, 167

\bibitem[Tremonti et al.(2007)]{Tremonti} Tremonti, C.~A., Moustakas, J. \& Diamond-Stanic, A.~M.\ 2007, \apj, 663, L77.



\bibitem[Voelk(1989)]{volk} Voelk, H.~J.\ 1989, \aap, 218, 67


\bibitem[Wright et al.(2010)]{wise} Wright, E.~L., Eisenhardt, P.~R.~M., Mainzer, A.~K., et al.\ 2010, The Astronomical Journal, 140, 1868


\bibitem[Yun et al.(2001)]{Yun} Yun, M.~S., Reddy, N.~A. \& Condon, J.~J.\ 2001, \apj, 554, 803.




}



\end{thebibliography}
\end{document}